# Layerwise Geo-Distributed Computing between Cloud and IoT

Satoshi Kamo, Yiqiang Sheng
Chinese Academy of Sciences, Beijing, China
University of Chinese Academy of Sciences, Beijing, China
Graduate School of Engineering, Osaka University, Osaka, Japan
Department of Communications and Integrated Systems, Tokyo Institute of Technology, Tokyo, Japan
E-mail: shengyq@dsp.ac.cn

*Abstract*—In this paper, we propose a novel architecture for a deep learning system, named *k*-degree layer-wise network, to realize efficient geo-distributed computing between Cloud and Internet of Things (IoT). The geo-distributed computing extends Cloud to the geographical verge of the network in the neighbor of IoT. The basic ideas of the proposal include a *k*-degree constraint and a layer-wise constraint. The *k*-degree constraint is defined such that the degree of each vertex on the $h^{th}$ layer is exactly $k^{(h)}$ to extend the existing deep belief networks and control the communication cost. The layer-wise constraint is defined such that the layer-wise degrees are monotonically decreasing in positive direction to gradually reduce the dimension of data. We prove the *k*-degree layer-wise network is sparse, while a typical deep neural network is dense. In an evaluation on the *M*-distributed MNIST database, the proposal is superior to a state-of-the-art model in terms of communication cost and learning time with scalability.

## 1. Introduction

With the rapid progress of communication, various state-of-the-art techniques such as Internet of Things (IoT) [1][2], deep learning [3][4][5], cloud computing and smart devices are increasingly emerged for enhancing our physical, social and virtual world towards an integrated world of big data. Efficient data processing based on everything connected to Internet has become a crucial requirement for many industrial applications . It is not only urgent but also necessary to make innovations through interdisciplinary researches on big data processing.

Cloud computing [6] partly meets the challenge of big data by collecting a number of computational resources together for data storage and data processing [7][8][9]. However, ever-growing requirements and techniques such as mobility-driven distributed event processing [10] have been making cloud computing inefficient due to expensive communication. The cloud-based real-time services are expensive because the distances of communication between cloud and IoTs are geographically far.

Geo-distributed computing such as on-site service [11] and fog [12][13] extends cloud to the verge of the Internet to geographically improve the quality of services in many scenarios. Since the computational resources of geo-distributed computing is in the neighbor of IoT, the communication cost is reduced. It makes easy to provide real-time services for end-users, but it leads to extra challenges to improve the computational performance of geo-distributed data processing.

Deep learning systems which aim at discovering higher levels of representations with more abstract concepts have achieved impressive performance in many applications [3][4][5] such as image recognition, speech recognition, multimedia data processing, information retrieval, human motion modeling and so on. However, big challenges such as extending a deep learning system to a large-scale model for efficient geo-distributed computing still lie ahead.

The main contributions of this research are as the following. (1) We propose a novel architecture for a deep learning system, named *k*-degree layer-wise network, to realize efficient geo-distributed computing between Cloud and IoT. (2) We define a general error function to evaluate training error, reconstruction error as well as regularization terms for generalization, and we investigate the proposed model with a *k*-degree constraint and a layer-wise constraint. (3) We prove the *k*-degree layer-wise network is sparse, while a typical deep neural network is dense. (4) Evaluation shows the proposal is superior to a state-of-the-art model in terms of the communication cost (more than 5X reduction) and the learning time (more than 3X speedup). Evaluation also shows the cost of the above improvement is the slower convergence speed with respect to data size.

The rest of this paper is organized as follows. Section 2 provides the related works. Section 3 is the formulation of learning model. Section 4 proposes the *k*-degree layer-wise network. Section 5 proves the sparsity of the proposed network. Section 6 shows the procedure to construct the proposed network. Section 7 shows the evaluation of the proposal in comparison with a state-of-the-art model of deep learning. Finally, Section 8 concludes this research.

## 2. Related Works

I. J. Lee [14] proposed the architecture of distributed processing which enables big data processing on the road traffic data and its related information analysis. It was a framework of road traffic collision using distributed complex event processing, and the massive real-time traffic data came from varying sources, such as social sites, mobile phone, GPS signals, and so on.

R. Tudoran et al [15] proposed a data management system on Cloud running across geographically distributed sites. The environment-aware solution monitored and modeled the global cloud infrastructure, and offered predictable data handling performance for communication cost and time. It provided the applications with the possibility to set a trade-off and optimized the communication strategy.

P. Bellavista et al [16] investigated the suitability of exploiting indications about differentiated priorities of stream processing tasks to enable application-specific resource of scheduling for distributed stream processing systems (DSPS). The priority-based resource scheduling was implemented on a stream processing application of vehicular traffic analysis by allowing application developers to augment graphs with priority meta-data and by introducing an extensible set of priority schema to be automatically handled by the extended DSPS.

R. Y. Shtykh and T. Suzuki [17] implemented distributed data processing in Onix which is a distributed data stream processing platform developed at CyberAgent and presented some scalability evaluation results.

K. Zhang and X. Chen [18] investigated a distributed deep learning model for restricted Boltzmann machines and back-propagation algorithm using MapReduce, a popular parallel programming model. As one of the state-of-the-art models of deep learning, deep belief net was trained in a distributed way by stacking a series of distributed restricted Boltzmann machines for pre-training and a distributed back-propagation for fine-tuning. Through validation on the benchmark data sets of various practical problems, the experimental results demonstrated the distributed deep belief net are amenable to large-scale data with a good performance in terms of accuracy and efficiency.

S. J. Rennie, P. Fousek and P. L. Dognin [19] presented distributed factorial hidden restricted Boltzmann machine. Speech and noise were modeled as independent restricted Boltzmann machines for robust speech recognition. The interaction between speech and noise was explicitly modeled to capture the combination to generate observed noisy speech features. An efficient algorithm which scales linearly with the number of hidden units was introduced to approximate the intractable inference which scales exponentially in the factorial hidden restricted Boltzmann machine. It had the advantage such that the representations of both speech and noise were distributed.

However, in the above researches, many serious problems of geo-distributed big data processing are not considered and evaluated. For example, the scalability of data size and the communication cost are two of the most serious problems to be investigated. Most of the state-of-the-art techniques such as deep belief network using MapReduce are limited to Cloud computing. Therefore, it is a big issue for efficient geo-distributed big data processing to meet the ever-growing requirements of Internet of Things and big data.

## 3. Formulation of Learning Model

Deep learning model is a class of neural networks with deep architecture [20][21]. It attracts wide attention due to the excellent performance [22][23] with open source tools and libraries [24][25]. The symbols of this paper are as follows. The data sets are marked as $D$ which includes $D_{train}$ as train sets, $D_{valid}$ as validation sets to select hyper-parameter and $D_{test}$ as test sets to evaluate the generalization error with a fair comparison between different models. The data set $D$ include a set of labeled data $Y = \{x_i, y_i\}$ and a set of unlabeled data $X = \{x_i\}$. Let $\theta$ be the set of all parameters. The objective is the generation error which measure the error of a model with respect to a generalized data set. A generalized error function of learning model is defined as the following.

$$E(\theta, D) = \sum_{i=1}^{l} E_i^t + \sum_{i=1}^{n} E_i^r + \sum_{i=1}^{q} \lambda_i L_i, \tag{1}$$

where $l$ is the number of labeled data, $n$ is the number of unlabeled and labeled data, $q$ is the number of regularization terms, $E_i^t$ is the training error with respect to each labeled datum $\{x_i, y_i\}$, $E_i^r$ is the reconstruction error with respect to each unlabeled and labeled datum, $\lambda_i$ is the weights of regularization terms, $L_1$ is the Lasso regularization term, $L_2$ is the Ridge regularization term, and so on. Let $y$ be output. Let $x$ be input. Let $F$ be the mapping of the model. A learning model is defined as the following.

$$y = F(x = x_i \mid \theta), \tag{2}$$

Let $G$ be the reversed mapping of $F$. The reversed model of reconstruction from the output to the input is defined as the following.

$$\hat{x} = G(y = y_i \mid \theta), \tag{3}$$

The training error with respect to each label $y_i$ is defined as the following.

$$E_i^{\ t} = \frac{1}{2}\|y - y_i\|^2,$$ (4)

Accordingly, the reconstruction error with respect to each input $x_i$ is defined as the following.

$$E_i^{\ r} = \frac{1}{2}\|\hat{x} - x_i\|^2,$$ (5)

The optimization procedure of parameter learning is as the following.

$$\theta^* = \arg\min_\theta E(\theta, D),$$ (6)

To evaluate the training error more efficiently, the learning process with respect to each labeled data $\{x_i, y_i\}$ is defined by minimizing the negative log-likelihood (*NLL*) as the following.

$$NLL = -\sum_i \log P(y = y_i \mid x_i, \theta),$$ (7)

As a state-of-the-art model of deep learning system, deep belief networks (DBN) [3][4][18] is a graphical model which learns to extract a deep representation of training data. DBN is stacked by restricted Boltzmann machines (RBMs) and pre-trained in a layer-wise greedy manner. The joint distribution between observed vector $x$ and the hidden layers $h_k$ is modeled as follows.

$$P(x, h_1, ..., h_l) = (\prod_{k=0}^{l-2} P(h_k \mid h_{k+1}))P(h_{l-1}, h_l),$$ (8)

where $x = h_0$, $P(h_k \mid h_{k+1})$ is a conditional distribution for the visible units conditioned on the hidden units of the RBM at layer $k$, and $P(h_{l-1}, h_l)$ is the visible-hidden joint distribution in the top-layer RBM.

However, more things have to be done to avoid over-fitting by early-stopping or regularization, when we are training our model from unknown data to do well on new samples. The regularized error function which penalizes certain parameters by adding the reconstruction error and regularization terms to the negative log-likelihood is as the following.

$$E(\theta, D) = NLL + \sum_i E_i^{\ r} + \sum_i \lambda_i L_i,$$ (9)

## 4. *k*-Degree Layer-Wise Network

In this research, a tree topology of two-level computer cluster with a core machine and $M$ verge machines is used. To match the tree topology, a hard constraint is designed as a class of non-flexible parameter setting. The sets of big data are divided into $M$ sub-sets according to the geographical sites, where $M$ is the number of verge machines. As a learning system, deep neural network with $H + 2$ layers, where $0 \le h \le H + 1$, is divided two levels by the layer of $h = H^*$ in the vertical direction of the system, as shown in Fig. 1.

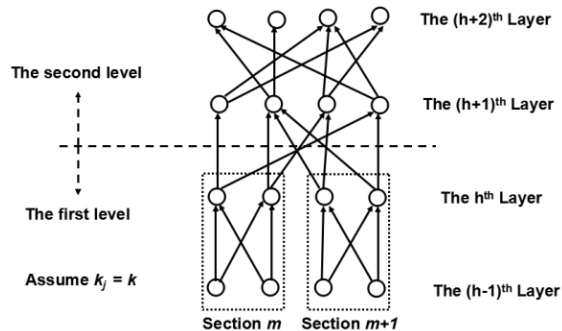

**Fig. 1** *k*-degree layer-wise network is proposed.

The layer of $h = 0$ is the input, the layer of $h = H^*$ is the partition, and the layer of $h = H + 1$ is the output. Accordingly, the first level of the system is from the layer of $h = 0$ to the layer of $h = H^*$, and the second level is from the layer of $h = H^* + 1$ to the layer of $h = H + 1$. Then, the first level of the system is divided $M$ sections in the horizontal direction of the system. The weights among different sections are zero as the hard constraint.

Let the width of the deep learning system with $H + 2$ layers be $N_w$, then the width of each section in the first level of the system is $d_m$, where $1 \le m \le M$, to ensure $N_w = \sum d_m$. Let the set of all neurons in the $m^{th}$ section of the first level be $S_m$. Let any neuron on the $(h-1)^{th}$ layer in the $m^{th}$ section be $S_{i(m),(h-1)} \in S_m$, where $1 \le h \le H^*$. Let the set of all neurons in the $o^{th}$ section of the first level be $O_m$. Let any neuron on the $l^{th}$ layer in the $o^{th}$ section be $S_{j(o),\ h} \in S_o$, where $m \ne o$. Then, the weights between $S_{i(m),(h-1)}$ and $S_{j(o),\ h}$ satisfy $w_{i(m),\ j(o),\ h} = 0$.

To improve the computational performance, the $k$-degree constraint is designed as a class of soft constraint which is an adaptive process of parameter learning. For a general non-hierarchical network, the $k$-degree constraint is defined such that the out-degrees of each neuron in positive direction are exactly $k$, where $k$ is a hyper parameter. The positive direction is from input to output.

Without the loss of generation, we assume all neurons are completely connected with adjustable weights. Let the number of all neurons be N. Let the network of neurons be connected with non-zero weights, and let the non-connected weight be zero. For the output of any neuron $x_j$, where $j = 1, 2, ..., N$, with the set of its inputs $X = \{x_i\}$, where $i = 1, 2, ..., k_j$, the following equation is satisfied.

$$x_j = f(\sum_{i=1}^{k_j} w_{ij} x_i + b_j),$$

(10)

where $f$ is the activation function, $b_j$ is the bias, and $w_{ij}$ is the weight from neuron $i$ to neuron $j$. For a general non-hierarchical network, the $k$-degree constraint is defined as the following.

$$k_j = k,$$

(11)

where $k$ is a constant of user-defined hyper parameter. For a hierarchical network, the $k$-degree constraint is defined as the following.

$$k_j^{(h)} = k^{(h)},$$

(12)

where $k_j^{(h)}$ is the number of the set of inputs $X = \{x_i\}$ for the output of neuron $x_j^{(h)}$ on the $h^{th}$ hidden layer, $k^{(h)}$ is a user-defined parameter for the $h^{th}$ hidden layer. The layer-wise constraint is defined such that the layer-wise degrees are monotonically decreasing in positive direction as the following.

$$\sum_j k_j^{(h+1)} \le \sum_j k_j^{(h)},$$

(13)

where $k_j^{(h+1)}$ is the number of the set of inputs $X = \{x_i\}$ for the output of neuron $x_j^{(h+1)}$ on the $(h+1)^{th}$ hidden layer. As a special case of (13), the layer-wise constraint could be the following.

$$k^{(h+1)} \le k^{(h)},$$

(14)

where $k^{(h+1)}$ is a user-defined parameter for the $(h+1)^{th}$ hidden layer. As a special case of the hierarchical network, the $k$-degree constraint for $0 < h < H$ could be defined as the following.

$$k_j^{(h)} = k,$$

(15)

## 5. Sparsity of $k$-Degree Layer-Wise Network

Here we prove the $k$-degree layer-wise network is sparse, while a typical deep neural network such as multi-layer perceptron or deep belief network is dense. For the sparsity of a network, let us start with a general network with $n$ vertices which can be represented as a graph $(V, E)$, where $E$ is the set of edges and $V$ is the set of vertices. For a directed simple graph, the density is defined as the following.

$$\rho(n) = \frac{|E|}{|V|(|V|-1)},$$

(16)

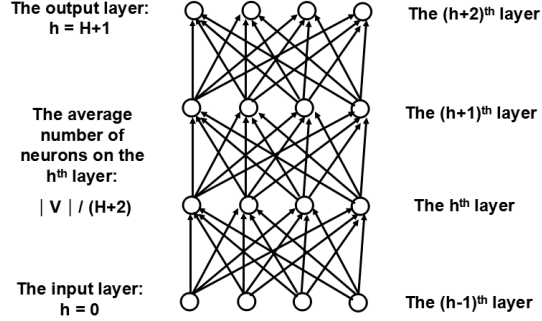

The output layer:
h = H+1

The average
number of
neurons on the $h^{th}$ layer:

| V | / (H+2)

The input layer:
h = 0

The $(h+2)^{th}$ layer

The $(h+1)^{th}$ layer

The $h^{th}$ layer

The $(h-1)^{th}$ layer

**Fig. 2** A typical deep neural network is dense.

A dense network is a network in which the number of edges is close to the maximal number of edges when $n$ approaches to infinite. The opposite, a network with a very small number of edges, is a sparse network when $n$ approaches to infinite. For a typical deep neural network [3][4][18], as shown in Fig. 2, the average number of neurons on the $h^{th}$ layer is as the following.

$$|V|^{(h)} \approx \frac{|V|}{H+2},$$

(17)

Since there are (H+1) groups of full connection between (H+2) layers in case of restricted Boltzmann machines, the number of edges is approximately satisfied as the following.

$$|E|_t \approx (\frac{|V|}{H+2})^2 (H+1),$$

(18)

Therefore, the typical deep neural network is dense because the density of network is approximately satisfied as the following.

$$\lim_{n \to \infty} \rho_t(n) \approx \frac{H+1}{(H+2)^2},$$

(19)

Although the average number of neurons on the $h^{th}$ layer of the $k$-degree layer-wise network is same as that of the typical deep neural network, there are (H+1) groups of $k$-degree connection between (H+2) layers in case of the $k$-degree layer-wise network. As shown in Fig. 1, the number of edges is approximately satisfied as the following.

$$|E|_k \approx \frac{k|V|}{H+2}(H+1),$$

(20)

Therefore, the $k$-degree layer-wise network which satisfies (11) is sparse because the density of network is satisfied as the following.

$$\lim_{n \to \infty} \rho_k(n) \approx \lim_{n \to \infty} \frac{k}{n-1} = 0,$$

(21)

## 6. Procedure

The $k$-degree layer-wise network is designed to construct a deep learning system by satisfying $k$-degree constraint and layer-wise constraint during the pre-training stage. It could be an unsupervised learning method which is using the set of unlabeled data. The detailed procedure is as the following.

Procedure (I): $k$-degree layer-wise network.

Input: A given model of deep neural network and the set of unlabeled data.

Output: A deep learning model with $k$-degree constraint.

Step 1: Coding from Input layer as $h$=0 to Output layer as $h$=$H$+1, where $H$ is the number of hidden layers. Let $h$ be -1.

Step 2: Initiating the parameters between the $h^{th}$ layer and the $(h+1)^{th}$ layer. Let $h$ be $h$+1.

Step 3: Adjusting the weights between the $h^{th}$ layer and the $(h+1)^{th}$ layer and the bias of the $(h+1)^{th}$ layer to minimize the cost function of the current two layers by using the set of unlabeled data.

Step 4: If the weight is smaller than the first threshold value, calculating $\Delta E_l^r$ which is the change of reconstruction error with and without the current connection. Deleting the current connection with the probability of $\min[1, \exp(-\Delta E_l^r/E_l^r)]$.

Step 5: Judging the out-degree of each neuron on the $h^{th}$ layer in the positive direction equals to $k^{(h)}$. If the answer is NO, returning to step 3. If the answer is YES, shifting to the next step.

Step 6: Judging the sum of all out-degrees of the $h^{th}$ layer in the positive direction is smaller than that of the $(h-1)^{th}$ layer or not. If the answer is NO, returning to step 3. If the answer is YES, shifting to the next step.

Step 7: Judging the current $h$ is larger than $H$ or not. If the answer is NO, returning to step 2. If the answer is YES, ending the procedure.

## 7. Evaluation

The evaluation of all models with learning algorithms was implemented by simulation using Python 2.7.8 [24] and Theano 0.6 [25] on a computer cluster which consisted of core machines and verge machines connected by 10Gbs switch. The core machine had 24 processors of 2.10GHz Intel Xeon E5-2620 CPU 32GB RAM and NVIDIA GPU Grid K2 8GB GDDR5. The verge machine had two processors of 2.50GHz Intel Core i7-4710MQ CPU 16GB RAM and NVIDIA GEFORCE GTX 760 GPU 2GB GDDR5.

To evaluate the communication cost, a computer cluster was described as a graph $(V, E, ED, EF)$, where $V$ is the vertex set $\{i\}$, $E$ is the edge set $\{e_{ij}\}$, $ED$ is the set of edge distance $\{ed_{ij}\}$ which is the geographical distance of communication between cloud and end-users, and $EF$ is the set of edge flow $\{ef_{ij}\}$ which is the data flow of model communication between cloud and end-users. The communication cost was defined as $CC = \sum_{ij} ed_{ij} ef_{ij}$. For two-level tree structure of computer cluster with a core machine and $M$ verge machines, the communication cost was simplified as $CC_{tree} = \sum_i ed_i ef_i$, where $ed_i$ is the geographical distance between the core machine and the verge machines, $ef_i$ is the data flow between the core machine and the verge machines using model communication, and $M$ is the number of the verge machines. In case of the model communication, the data flow $ef_{ij}$ of communication between cloud and Internet of Things equals to the number of parameters $N_{ij}(\theta)$.

For $M$-distributed MNIST database, each MNIST handwritten image was splitted into $M$ parts to allocate each part to one of verge machines, and each splitted data set is with 50, 000 samples for training to optimize the given parameters, 10, 000 samples for validation to select hyper-parameter and 10,000 samples for testing to evaluate the generalization error with a fair comparison between different models. All splitted images were with a fixed size of 784/$M$ pixels. Besides, the training set of MNIST with 50, 000 samples was repeated with X times to evaluate the scalability performance, where X = 2, 3, 4, 5, ..., 100 and X = 10, 100, 1000, 10000.

The main procedure to train deep learning models [1][2] is as the following.

Procedure (II): Parameter optimization process of deep learning system.

Input: The set of all parameters of a given model of deep neural network $\theta$ and the set of unlabeled data and labeled data $D$.

Output: A minimized cost function $E(\theta, D)$ with the optimized set of parameters $\theta$.

Step 1: Let $h$ be 1. Training the first layer ($h$) as a restricted Boltzmann machine which models the raw input $x = h_0$ as its visible layer ($h$-1) with given constraints.

Step 2: Using the above first layer ($h$) to obtain a representation of the input which is used as data for the second layer ($h$+1). This representation is chosen as being the mean activation values $P(h_1 = 1 \mid h_0)$ or samples of $P(h_1 \mid h_0)$.

Step 3: Training the second layer ($h$+1) as a restricted Boltzmann machine with the constrains and taking the transformed data as training examples for the visible layer of that restricted Boltzmann machine with the constrains.

Step 4: Let $h$ be $h$+1. Iterating Step 2 and Step 3 for the desired number of layers ($h$=$H$) with the constrains, each time propagating upward either samples or mean values.

Step 5: Fine-tuning all the parameters of this deep architecture with respect to the cost function $E(\theta, D)$ or its proxy as a training criterion.

For the hierarchical network, we select $M = 5$, $k^{(h)} = 50$ ($h < H$) and $k^{(H)} = 10$ to test the performance of the algorithms. The input layer ($L_0$) has 784 neurons. The output layer ($L_6$) has 10 neurons. The default setting of hidden layers for the tested model is as follows. The first hidden layer ($L_1$) has 580 neurons. The second hidden layer ($L_2$) has 450 neurons. The third hidden layer ($L_3$) has 310 neurons. The fourth hidden layer ($L_4$) has 160 neurons. The fifth hidden layer ($L_5$) has 75 neurons. $L_0$, $L_1$ and $L_2$ are in the verge machines, while $L_3$, $L_4$, $L_5$, and $L_6$ are in the core machine. For DBN, the layer-wise edges from the input layer to the output layer are 455504, 261580, 139950, 49910, 12160 and 825, respectively. For the proposal, the layer-wise edges from the input layer to the output layer are 39200, 29000, 22500, 15500, 8000 and 750, respectively. The regularization terms ($L_j$) include $L_1$ and $L_2$. To avoid the over-fitting of parameter learning, the weights ($\lambda_1$ and $\lambda_2$) of regularization terms are set to be small numbers. 0.00001 and

0.00009 are two empirical values determined by trial experiments. The remaining parameters are optimized during the same learning process of deep learning systems with different constraints.

We evaluated one of the state-of-the-art deep learning models named deep belief net (DBN) which was trained by stacking a series of restricted Boltzmann machines for pre-training and a back-propagation for fine-tuning in a distributed way to get the best performance in terms of error rate, running time and communication cost. The communication cost of the parameter optimization processes of distributed DBN is shown in Table 1. We observed that the communication is expensive with respect to the scaling of data size as a necessary and urgent problem. The learning time of the parameter optimization processes of distributed DBN is shown in Table 2. The learning time is gradually increased with respect to the scaling of data size. When the data size is 50, 000, the learning time is nearly 30 seconds. When the data size increases 1000 times, the learning time is nearly 8 hours.

**Table 1** The communication cost of distributed learning systems

| Data Size | Communication Cost (50KB) | | Improvement |
|---|---|---|---|
| (50K) | DBN | $k$-degree | (times) |
| 1 | 20 | 4 | 5.0 |
| 2 | 36 | 7 | 5.1 |
| 4 | 71 | 14 | 5.1 |
| 6 | 107 | 20 | 5.4 |
| 8 | 143 | 27 | 5.3 |
| 10 | 178 | 34 | 5.2 |
| 100 | 1784 | 340 | 5.2 |
| 1000 | 17838 | 3396 | 5.3 |
| 10000 | 178383 | 33962 | 5.3 |

**Table 2** The learning time of distributed learning systems

| Data Size | Learning Time (s) | | Speedup |
|---|---|---|---|
| (50K) | DBN | $k$-degree | (times) |
| 1 | 29 | 9 | 3.2 |
| 2 | 56 | 17 | 3.3 |
| 4 | 119 | 35 | 3.4 |
| 6 | 178 | 51 | 3.5 |
| 8 | 236 | 67 | 3.5 |
| 10 | 295 | 81 | 3.6 |
| 100 | 2941 | 797 | 3.7 |
| 1000 | 29514 | 7963 | 3.7 |
| 10000 | 294811 | 79446 | 3.7 |

We evaluated the $k$-degree layer-wise network which was a novel architecture for deep neural network with the given constraints in order to improve the performance of distributed data processing. The $k$-degree constraint is a soft constraint which is an adaptive process of parameter learning to ensure that the out-degrees of each neuron in positive direction are exactly $k$. The layer-wise constraint is also a soft constraint which is an adaptive process of parameter learning to ensure that the layer-wise degrees are monotonically decreasing in positive direction. As shown in procedure (I), the $k$-degree layer-wise network is used to construct an efficient deep learning system to satisfy all soft constraints. The communication cost of the parameter optimization processes of the proposal is shown in Table 1. As a result, the communication cost is from 5.0 to 5.4 times better than that of DBN. The learning time of the parameter optimization processes of the proposal is shown in Table 2. As a result, the learning time is from 3.2 to 3.7 times shorter than that of DBN. Although the improvement is not so stable, both communication cost and learning time increase almost linearly.

**Table 3** The speed of convergence with respect to data size as the cost of improvements (Basic error rate: 10%; Basic data size: 50K)

| Data Size | Error Rate (%) | | Convergence Speed (%) | | Relative Cost (%) |
|---|---|---|---|---|---|
| (50K) | DBN | $k$-degree | DBN | $k$-degree | (%) |
| 1 | 4.92 | 7.71 | 2.54 | 1.14 | 54.95 |
| 2 | 4.46 | 7.09 | 1.85 | 0.97 | 47.48 |
| 3 | 4.09 | 6.61 | 1.48 | 0.85 | 42.67 |
| 4 | 3.77 | 6.23 | 1.25 | 0.75 | 39.45 |
| 5 | 3.53 | 5.92 | 1.08 | 0.68 | 37.02 |

| 6 | 3.38 | 5.68 | 0.95 | 0.62 | 34.73 |
| 7 | 3.30 | 5.45 | 0.84 | 0.57 | 32.17 |
| 8 | 3.15 | 5.20 | 0.76 | 0.53 | 29.90 |
| 9 | 3.07 | 4.96 | 0.69 | 0.50 | 27.35 |
| 10 | 2.95 | 4.78 | 0.64 | 0.47 | 25.98 |

The cost of the above improvement is the slower convergent speed with respect to data size, but the total learning time is actually improved because the proposed network dramatically reduces the learning time with the similar data size by deleting unnecessary connection under the constraints. Let the convergence speed denote the improvement of error rates divided by the data size as the following.

$$CS_i = \frac{N_0(d)}{N_0(d) + N_i(d)}(E_0 - E_i), \quad (22)$$

where $CS_i$ is the current convergence speed, $E_i$ is the current error rate, $E_0$ is the basic error rate, $N_i(d)$ is the current data size and $N_0(d)$ is the basic data size.

Let the relative cost denote the relative degradation of the convergence speed. The detail of the error rate, the convergence speed and the relative cost with respect to data size is shown in Table 3. The basic error rate to calculate the convergence speed and the relative cost is 10%. The error rate of the parameter optimization processes of distributed DBN and the proposal is shown in Fig. 3. The accuracy or error rate is gradually improved with respect to the data size of learning. When the data size is small, the error rate of the proposal is not as good as that of a state-of-the-art model of deep learning. However, this does not matter, because we care more about the final result of parameter learning and the accuracy is gradually improved when the data size increases. When the data size is nearly 100, the error rate of the proposal is as good as that of a state-of-the-art model of deep learning.

That is to say, the improvement of communication cost and learning time comes with no degradation of accuracy, when the number of data sizes are large enough. The exact reason in theory is still unknown, but a part of reason should be as the following. The distribution of near-optimal solutions of the DBN system is much denser than that of the proposed system, while the quality of optimal solutions of both systems is on the same level.

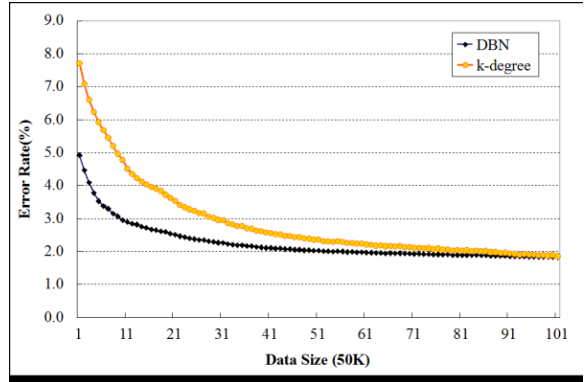

**Fig. 3** Comparison of the parameter optimization processes of two different learning systems.

## 8. Conclusion and Future Work

In this paper, a novel architecture for a deep learning system, named *k*-degree layer-wise network, was proposed to realize big data processing via efficient geo-distributed computing between cloud and IoT. We proved that the *k*-degree layer-wise network is sparse, while a typical deep neural network is dense. We investigated the *k*-degree constraint and the layer-wise constraint. Evaluation shows the considerable improvement of performance. For geo-distributed big data processing, the communication cost of the proposal is at least 5 times smaller than that of the state-of-the-art model of deep learning with shorter learning time and without degradation of accuracy. The cost of the above improvement is the slower convergent speed with respect to data size, but the final learning time is actually improved (more than 3.2X) for big data. Though the division number M is not explicitly discussed this time, it is an important factor which impacts the final performance of data processing. It is necessary to investigate the factor with scalability in future work [26].

**Acknowledgment**

This work is partly supported by Special Fund for Strategic Pilot Technology of Chinese Academy of Sciences under Grant No. XDA06040501. The authors would like to thank Prof. X. Zeng, Dr. L. Wang and Dr. W. Qi at Chinese Academy of Sciences, thank

Prof. A. Takahashi and Prof. S. Ueno at Tokyo Institute of Technology, and thank the anonymous reviewers for their valuable comments.